# Phase-separated high-temperature-annealed (Ga,Mn)As:

# A negative charge-transfer-energy material

M. Moreno[a,*] and K. H. Ploog[b]

[a] *Instituto de Ciencia de Materiales de Madrid (CSIC), Cantoblanco, E-28049 Madrid, Spain*

[b] *Paul-Drude-Institut für Festkörperelektronik, Hausvogteiplatz 5-7, D-10117 Berlin, Germany*

(Date: August 4, 2011)

## ABSTRACT

The approximate location in the Zaanen-Sawatzky-Allen diagram of the phase-separated (Ga,Mn)As material, consisting of MnAs nanoclusters embedded in GaAs, is determined on the basis of configuration-interaction (CI) cluster-model analysis of their Mn $2p$ core-level photoemission. The composite material is found to belong to the special class of materials with *negative* charge-transfer energy ($\Delta$). As such, its metallic or insulating/semiconducting behavior depends on the strength of the *p-d* hybridization (affected by strain) relative to the (size-dependent) *p*-bandwidth. Whereas internal strain in the embedded clusters counteracts gap opening, a metal-to-semiconductor transition is expected to occur for decreasing cluster size, associated to the opening of a small gap of *p-p* type (covalent gap). The electronic properties of homogeneous and phase-separated (Ga,Mn)As materials are analyzed, with emphasis on the nature of their metal-insulator transitions.

---

[*] E-mail: mmoreno@icmm.csic.es

# I – INTRODUCTION

The temperature-composition ($T$-$x$) phase diagram of the Ga$_{1-x}$Mn$_x$As alloy exhibits a miscibility gap, within which phase separation, into GaAs and MnAs binary phases, occurs at thermal equilibrium.[1] The miscibility gap arises from the thermodynamic tendency of Mn impurities to cluster, driven by the formation of energy lowering Mn–As–Mn bonds.[2] Thin films of single-phase homogeneous Ga$_{1-x}$Mn$_x$As can be fabricated by low-temperature molecular-beam epitaxy on GaAs substrates and rapid quenching from growth down to room temperature. On the other hand, phase separation can be promoted by annealing homogeneous Ga$_{1-x}$Mn$_x$As at temperatures above the growth temperature. Mild annealing induces spinodal decomposition: an inhomogeneous Mn distribution across the crystal, with regions of high and low Mn density, the zincblende structure being maintained.[3] High-temperature annealing induces nucleation of small hexagonal-MnAs clusters within the zincblende GaAs matrix.[3,4]

The GaAs:MnAs composite material, consisting of hexagonal-MnAs nanoclusters embedded in zincblende GaAs, has properties intermediate between those of bulk MnAs and those of homogeneous Ga$_{1-x}$Mn$_x$As. MnAs is a concentrated magnetic material, which crystallizes in the hexagonal NiAs-type structure ($D_{3d}$ point group) and exhibits robust ferromagnetism as well as metallic conductivity at room temperature. In Ga$_{1-x}$Mn$_x$As, diluted concentrations ($x<\sim0.1$) of Mn atoms randomly substitute for Ga atoms in the zincblende GaAs structure. Ga$_{1-x}$Mn$_x$As is ferromagnetic only at temperatures far below room temperature ($T_C<190$ K), and its conduction properties indicate the proximity to a metal-insulator transition. MnAs nanoclusters embedded in GaAs are ferromagnetic at room temperature.[3,5] Sufficiently large clusters are metallic, but a metal-insulator (metal-semiconductor) transition is expected to occur for decreasing cluster size. Small MnAs nanocrystals embedded in GaAs have been recently shown to exhibit remarkably long spin relaxation times.[6]

The mechanism of the metal-insulator transition in $Ga_{1-x}Mn_xAs$ is not clear yet, neither its ferromagnetism mechanism. It is widely accepted that Mn acts as an acceptor in GaAs, and that the Mn-derived holes mediate the ferromagnetic coupling between the local Mn moments. However, the degree of localization of these holes is an intensively discussed issue. There is considerable debate about whether the Mn-derived impurity band, where holes reside, is merged with the GaAs valence band or decoupled from it.[7-9] According to Mott theory[10] for the metal-insulator transition (MIT), considered within the Zener model of $Ga_{1-x}Mn_xAs$ ferromagnetism,[11] Mn is expected to initially form an impurity band that eventually merges with the GaAs valence band at high enough Mn impurity densities. However, while the Mott criterion is valid for the MIT of semiconductors doped with high concentrations of shallow, nonmagnetic donors and acceptors, which have hydrogenic states of extended nature, its applicability to $Ga_{1-x}Mn_xAs$ raises serious doubts. First-principles LDA+$U$ calculations have shown[12] that the picture of a delocalized (host-like) hole is invalid for $Ga_{1-x}Mn_xAs$, since for the $U$ that leads to agreement with XPS results,[13] the hole is still localized to some extent. Since Mn is not a conventional acceptor in GaAs, the generated holes not being hydrogenic-like, Mott theory is not necessarily applicable, and the metal-insulator transition is not necessarily controlled (directly) by the Mn concentration. Several theoretical and experimental investigations have indicated that the Mn-derived impurity band remains detached from the valence band at high values of the Mn concentration.[14-17] Recent experimental results have provided strong evidence for the holes (mediating ferromagnetism) residing in states generated by the Mn impurities and not in the GaAs valence band.[7] The mechanism of the metal-insulator transition in $Ga_{1-x}Mn_xAs$ remains to be clarified.

An understanding of how electronic properties evolve on going from single-phase homogeneous $Ga_{1-x}Mn_xAs$ to phase-separated GaAs:MnAs materials is desirable. It is also important to fully understand the size-induced changes in the electronic structure of MnAs.

We recently reported on a comparative photoemission study of the electronic structure of MnAs, as thick film and as nanoclusters embedded in GaAs.[18] Here, we complementary report on the approximate location of MnAs nanoclusters in the Zaanen-Sawatzky-Allen (ZSA) phase diagram,[19,20] determined on the basis of configuration-interaction (CI) cluster-model analysis of its Mn 2$p$ photoemission. We make a rational connection between the electronic properties of homogeneous $Ga_{1-x}Mn_xAs$ and phase-separated GaAs:MnAs materials, and we get insight into the nature of the metal-insulator transitions in these materials.

## II - EXPERIMENTAL DETAILS

We grew a $Ga_{0.916}Mn_{0.084}As$ thin film on a heavily $n$-type Si-doped GaAs (001) substrate by low-temperature molecular-beam epitaxy. After growth, the sample was in-situ annealed, first at 247 °C for 5 min, and then at 640 °C for 10 min. The second annealing step leads to the precipitation of small hexagonal-MnAs nanoclusters embedded in a GaAs matrix. After annealing, the sample was cooled down well below room temperature, and exposed to an $As_4$ flux in order to deposit a protective arsenic coating. The arsenic-coated sample was briefly exposed to air during transfer from the growth to the analysis chamber. Photoelectron emission spectroscopic (PES) analyses were carried out on the soft-X-ray undulator beamline UE56/2-PGM-2 of the BESSY II facility using an ESCALAB MkII analyzer. The arsenic coating was desorbed by heating the sample up to ~370 °C.

## III - RESULTS

Figure 1 shows a PES spectrum of the nanoclusters Mn 2$p$ core level, recorded at room temperature, with synchrotron light (hν = 750 eV) incident 30º off normal and photoelectrons collected in the direction normal to the sample surface in an angle-integrated mode. The nanoclusters spectrum consists of a $2p_{3/2}$-$2p_{1/2}$ spin-orbit split doublet, with a complex structure. Each component of the doublet consists of a leading signal (m) and a *correlation* satellite structure at 3-5 eV higher binding energy. In addition, energy losses due to plasmon excitation contribute to the nanoclusters spectrum, at roughly 19 eV higher binding energy than the leading signals. The presence of correlation satellite structures in the Mn $2p$ spectrum evidences the poor screening ability of valence electrons in the nanoscaled material. The Mn $3d$ charge fluctuates, such that several photoemission final states contribute to the Mn $2p$ spectrum.

In order to obtain the characteristic parameters of the nanoclusters electronic structure, we have analyzed the lineshape of their Mn $2p$ PES spectrum on the basis of a CI cluster model.[21] In the cluster approximation, we consider an idealized structure, a $MnAs_6$ cluster, consisting of just a central Mn cation surrounded by six As anions, arranged in a NiAs-type MnAs structure with the lattice parameters measured[22,23] for hexagonal-MnAs nanoclusters embedded in GaAs. Three parameters characterize the electronic structure of the $MnAs_6$ cluster: the on-site $3d$-$3d$ Coulomb repulsion energy $U$, the energy required to transfer one electron from a ligand As $4p$ orbital to a Mn $3d$ orbital (charge-transfer energy) $\Delta$, and the Slater-Koster parameter $(pd\sigma)$ accounting for the strength of the As $4p$-Mn $3d$ hybridization. In the limit of 100% ionicity for the Mn-As bond, the Mn cation donates three electrons to its As neighbors, becoming $Mn^{3+}$ charged ($d^4$ configuration). The wave functions are described, in the CI picture, as linear combinations of the "ionic" configuration, $d^4$, and "charge-transfer" configurations, $d^{4+m}\underline{L}^m$, where one or more (*m*) electrons are transferred from As $4p$ orbitals to the Mn $3d$ levels ($\underline{L}$ stands for a hole in a As $4p$ orbital).

In absence of *p-d* hybridization, the ionic $d^4$ and charge-transfer $d^{4+m}\underline{L}^m$ states are eigenstates. The diagonal matrix elements of the corresponding Hamiltonian account[21] for the charge-transfer energy, $\Delta \equiv E(d^{n+1}\underline{L}) - E(d^n)$, and for the Coulomb repulsion energy, $U \equiv E(d^{n-1}) + E(d^{n+1}) - 2E(d^n)$. The photoemission final states differ from the initial states by the Coulomb attraction $Q$ between the Mn $2p$ core hole and the Mn $3d$ valence electrons, which pulls down the charge-transferred states relative to the ionic state.[24] Due to anisotropic intra-atomic Coulomb and exchange interactions, the ionic $d^4$ and charge-transfer $d^{4+m}L^m$ states are split into multiplet terms. Each multiplet term is split from the center of gravity of the multiplet, $E(d^n)$, by a Coulomb-exchange interaction energy correction, $\Delta E(d^n: {}^{2S+1}\Gamma)$, which is expressed[21] in terms of the Racah parameters $B$ and $C$, here fixed at the values of the free $Mn^{3+}$ ion $B=0.120$ eV, $C=0.552$ eV (Ref. 21). We account for crystal-field effects and for the anisotropic $d$-$d$ Coulomb-exchange interactions by extending the basis set to distinguish between each individual hexagonal-MnAs orbital symmetry (Ref. 25) and spin state: $a_1^T\uparrow, a_1^T\downarrow, e^T\uparrow, e^T\downarrow, e\uparrow,$ and $e\downarrow$. The high-spin ionic state $d^4 = (a_1^T\uparrow)^1(e^T\uparrow)^2(e\uparrow)^1$ is taken as reference level. While $\Delta$ and $U$ are multiplet-averaged quantities, defined with respect to the centers of gravity of the multiplets, the charge-transfer energy and the Coulomb repulsion energy can also be defined with respect to the lowest terms of the multiplets. The effective parameters thus defined are given by $\Delta_{\text{eff}}=\Delta+\Delta E_n-\Delta E_{n+1}$ and $U_{\text{eff}}=U+2\Delta E_n-\Delta E_{n-1}-\Delta E_{n+1}$, where $-\Delta E_n$ denotes the $\Delta E(d^n: {}^{2S+1}\Gamma)$ energy correction for the lowest term of the $d^n$ multiplet.[26] For high-spin $Mn^{3+}$ ($d^4$) compounds under $D_{3d}$ symmetry, $\Delta_{\text{eff}} = \Delta + 4(\frac{14}{9}B - \frac{7}{9}C) - 14B$ and $U_{\text{eff}} = U + (\frac{14}{9}B - \frac{7}{9}C) - \frac{7}{2}B$. Considering the $B$ and $C$ values corresponding to the free $Mn^{3+}$ ion, they are given by $\Delta_{\text{eff}} = \Delta - 2.56$ and $U_{\text{eff}} = U - 0.64$ (in eV units).

Hybridization between the Mn $3d$ and As $4p$ orbitals is treated, in the CI picture, as a perturbation.[21] It is included in off-diagonal (mixing) matrix elements of the Hamiltonian, via one-electron transfer integrals $T \equiv \langle d|H|L\rangle$, expressed in terms of the Slater-Koster parameters $(pd\sigma)$ and $(pd\pi)$. To account for the anisotropic $p$-$d$ hybridization in MnAs, we consider different transfer integrals for the orbitals with $a_1^T, e^T$, and $e$ symmetries. Off-diagonal matrix elements of the final-state Hamiltonian are assumed to be the same as those of the initial-state Hamiltonian.

The photoemission spectrum is given, in the sudden approximation, by the superposition of the δ-lines corresponding to the different final-state eigenenergies.[21] To simulate the experimental spectrum, we broaden the theoretical photoemission δ-lines by Gaussian/Doniach-Sunjic convoluted functions, representing the experimental resolution and intrinsic photoemission lineshape. We varied the parameters $U$, $\Delta$, and $(pd\sigma)$ to obtain the best fit to the experimental data, shown in Fig. 1. The best-fit values are listed in Table I. The CI analysis reveals that the leading (m) peak and the correlation satellite structure correspond to transitions to different photoemission final states, mixing $\underline{2p}d^6\underline{L}^2$, $\underline{2p}d^7\underline{L}^3$, and $\underline{2p}d^5\underline{L}$ configurations mainly, where $\underline{2p}$ stands for a Mn $2p$ core hole. The correlation satellite structure is revealed to consist of several contributions, the most intense ones being "s1" and "s2", at 5.2 and 3.4 eV higher binding energies than the leading "m" line, respectively (Fig. 1). Whereas the leading (m) photoemission signal is found to be mainly contributed by the $\underline{2p}d^6\underline{L}^2 = \underline{2p}(a_1^T \uparrow)^1(e^T \uparrow)^2(e \uparrow)^2(e \downarrow)^1(\underline{L}_e \downarrow)^1(\underline{L}_e \uparrow)^1$ configuration, the correlation satellites (s1 and s2) are found to be mainly contributed by the $\underline{2p}d^5\underline{L} = \underline{2p}(a_1^T \uparrow)^1(e^T \uparrow)^2(e \uparrow)^2(\underline{L}_e \downarrow)^1$ configuration.

## IV - DISCUSSION

### A. MnAs and GaAs:MnAs in the ZSA phase diagram

According to DFT electronic-structure calculations,[27-29] bulk MnAs is a metal with hybrid character: the spin-up channel is ($pd$ metal)-like and the spin-down channel is ($d$ metal)-like. The on-site $d$-$d$ Coulomb repulsion in bulk MnAs is theoretically estimated[29] to be low ($U \approx 0.98$ eV), consistent with the weakness/absence of correlation satellites in Mn $2p$ photoemission from thick MnAs films.[18] In contrast, Mn $2p$ photoemission from MnAs nanoclusters embedded in GaAs exhibits pronounced correlation satellites (Fig. 1), evidencing the notable enhancement of the on-site $d$-$d$ Coulomb repulsion that occurs upon nanoscaling. The CI analysis of Mn $2p$ photoemission reveals that the Coulomb repulsion energy for the nanoclusters is about $U= 3.6$ eV ($U_{eff} = 3.0$ eV). Most important, it reveals that MnAs nanoclusters embedded in GaAs belong to the special class of materials with *negative* charge-transfer energy ($\Delta = -2.1$ eV and $\Delta_{eff} = -4.7$ eV).

Figure 2 shows the energy-level diagram for the nanoclusters in the MnAs$_6$ cluster model. In absence of $p$-$d$ hybridization (left-side diagram), the $d^5\underline{L}$ state is the ground state, consistent with the negative value of $\Delta$ [$E(d^5\underline{L}) \equiv E(d^4) + \Delta$]. The $d^5\underline{L}$ and $d^6\underline{L}^2$ charge-transfer states are seen to be stabilized with respect to the $d^4$ ionic state. Switching on $p$-$d$ hybridization (right-side diagram), the $d^4$ and $d^{4+m}\underline{L}^m$ multiplets split off and configurations mix up. The ground state then corresponds to a strongly mixed state, consisting of $d^5\underline{L}$ (66% weight), $d^6\underline{L}^2$ (25% weight), and $d^4$ (4% weight) configurations. It is largely dominated (60% weight) by the $d^5\underline{L} = (a_1^T \uparrow)^1 (e^T \uparrow)^2 (e \uparrow)^2 (\underline{L}_e \downarrow)^1$ configuration, which corresponds to half $d$-orbital filling, maximum local Mn moment, and one spin-up hole in As $4p$ orbitals with $e$ symmetry. The $d$-electron count is found to be 5.22, such that Mn cations have +1.78 effective valence.

### B. Ga$_{1-x}$Mn$_x$As in the ZSA phase diagram

CI cluster-model analyses were previously carried out for Mn $2p$ photoemission from a Ga$_{0.93}$Mn$_{0.07}$As diluted magnetic semiconductor (Ref. 30). Two different values were reported for the charge-transfer energy $\Delta$. Whereas a positive value ($\Delta = +1.5$ eV) was found considering $Mn^{2+}$ oxidation state as starting point of the CI analysis, a negative value ($\Delta = -1.5$ eV) was found considering $Mn^{3+}$ oxidation state. The (dominant) Mn configuration was found to be $d^5$ in both cases. The notable discrepancy on the value of $\Delta$ is of the highest relevance. The sign of $\Delta$ is different, such that the CI analysis[30] identifies $Ga_{1-x}Mn_xAs$ as either a negative-$\Delta$ material or a charge-transfer material ($\Delta > 0$) depending on the oxidation state considered. These are distinct classes of materials, with sharply contrasting properties. We note, for instance, that the effect on the metal-insulator transition, of varying the *p-d* hybridization, is *opposite* in the charge-transfer ($\Delta > 0$) and negative-$\Delta$ regimes.[31] The authors of this early work[30] left the issue of the location of $Ga_{1-x}Mn_xAs$ in the ZSA diagram unresolved.

Each transition-metal (TM) compound has a well-defined location in the ZSA diagram, which identifies the nature of the states crossing the Fermi energy, in the case of metals, or the type of gap, in the case of insulators. CI cluster-model analyses of photoemission spectra allow to estimate the parameters ($U$, $\Delta$, and $pd\sigma$) that roughly locate TM compounds in the ($U$-$\Delta$) ZSA phase diagram. However, for the parameters to have universal meaning, allowing comparison to other materials and proper location in the ZSA diagram, the CI analysis must be carried out considering the proper nominal valence (oxidation state) for the TM cation, which corresponds to the limit of 100% ionicity for the TM-ligand bond, that is, to the hypothetical situation where *p-d* hybridization is "switched off" (i.e., covalency is set to zero) and the ligands (anions) complete their outer *p* shell taking the necessary electrons from the cations. The nominal valence must not be assimilated to the "effective" valence, the later being the actual valence, accounting for *p-d* hybridization, i.e., for covalency of the TM-ligand bond. The oxidation state of manganese in compounds depends on the compound considered. Common oxidation states are $Mn^{2+}$, $Mn^{3+}$, and $Mn^{4+}$. To name a few examples, the oxidation state is $Mn^{2+}$ in binary chalcogenides (group-VI compounds, e.g., MnSe) as well as in II-VI diluted magnetic semiconductors (e.g., $Cd_{1-x}Mn_xSe$).[32-34] The oxidation state is $Mn^{3+}$ in $LaMnO_3$ (Refs. 35,36), and $Mn^{4+}$ in $SrMnO_3$ (Refs. 36,37).

In pnictides (group-V compounds, e.g., MnAs), as well as in III-V diluted magnetic semiconductors (e.g., $Ga_{1-x}Mn_xAs$), the oxidation state is $Mn^{3+}$, since the group-V ligands require 3 electrons to completely fill their outer $p$ shell. $Mn^{2+}$ is not the proper oxidation state (*nominal* valence) to be considered in CI analyses of $Ga_{1-x}Mn_xAs$ photoemission spectra, since the limit of 100% ionicity is then not properly set; the condition of charge neutrality is not fulfilled, an unphysical negative charge ($x-$) being introduced $\left[(Ga^{3+}_{1-x}Mn^{2+}_{x}As^{3-})^{x-}\right]$. If $Mn^{2+}$ oxidation state is considered in the CI analysis,[30] the $\Delta$ value obtained does not refer to a universal scale; it does not properly locate $Ga_{1-x}Mn_xAs$ in the ($U$-$\Delta$) ZSA phase diagram. In order to correctly set the limit of 100% ionicity in CI analyses of $Ga_{1-x}Mn_xAs$ photoemission, the $Mn^{3+}$ oxidation state has to be considered. The $\Delta$ value thus obtained does refer to a universal scale. Consequently, $Ga_{1-x}Mn_xAs$ belongs to the class of negative-$\Delta$ materials. The ($Mn^{3+}$) CI cluster-model analysis[30] reveals that the $Ga_{1-x}Mn_xAs$ ground state is dominated by the $d^5\underline{L}$ charge-transfer configuration. The ($Mn^{3+}$) CI analysis explains the origin of the hole ($\underline{L}$) mediating ferromagnetism, making no assumption on its localized or free nature. The later depends on band structure effects that go beyond the cluster approximation employed in the analysis. The ($Mn^{3+}$) CI analysis is compatible with either a localized, a partially localized, or a free ($\underline{L}$) hole.

A *negative* $\Delta$ value can a priori be expected for $Ga_{1-x}Mn_xAs$, GaAs:MnAs, as well as for MnAs, considering the chemical trend that Mn compounds follow (Table I). The charge-transfer energy $\Delta$ is lower for Mn compounds with group-VI ligands (e.g., MnS, MnSe, MnTe) than for Mn compounds with group-VII ligands (e.g., $MnF_2$, $MnCl_2$, $MnBr_2$), since $\Delta$ decreases with decreasing electronegativity of the ligand. Similarly, the charge transfer energy is expected to be lower for Mn compounds with group-V ligands (e.g., MnAs, $Ga_{1-x}Mn_xAs$) than for Mn compounds with group-VI ligands. Considering that $\Delta$ is already very low for MnTe ($\Delta=0$ according to Ref. 33), the chemical trend indicates that $Ga_{1-x}Mn_xAs$, GaAs:MnAs, and MnAs should be negative-$\Delta$ materials. The charge-transfer energy $\Delta$ is found to be somewhat lower (more negative) for embedded MnAs nanoclusters with octahedral coordination, $\Delta = -2.1$ eV, than for $Ga_{0.93}Mn_{0.07}As$ with tetrahedral coordination, $\Delta = -1.5$ eV (Table I). A similar trend is observed for Mn compounds with group-VI ligands: the charge-transfer energy $\Delta$ is lower for the binary compound with octahedral coordination (e.g., MnTe, $\Delta = 0$), than for diluted magnetic semiconductors with tetrahedral coordination (e.g., $Cd_{1-x}Mn_xTe$, $\Delta = 2.0$ eV).

### C. Metal-insulator transition in GaAs:MnAs

The identification of homogeneous and phase-separated $Ga_{1-x}Mn_xAs$ as negative-$\Delta$ materials provides important clues on the mechanisms of metal-insulator transitions occurring in this material system. Whereas metal-insulator transitions in the charge-transfer and Mott regimes are controlled by the values of $\Delta$ and $U$, respectively, the metal-insulator transition in the negative-$\Delta$ regime is controlled by the strength of the *effective p-d* hybridization relative to the *p*-bandwidth.[20,31] For negative-$\Delta$ materials, a gap of *p-p* type opens if the effective hybridization is sufficiently strong and/or the *p*-bandwidth is sufficiently narrow. The effective *p-d* hybridization increases with decreasing TM-ligand bond length, but it weakens with *extended* coupling of different TM-ligand units.[20,31] Extended coupling is maximized when the TM-ligand-TM bond angle is close to 180°, and minimized when the angle is close to 90° (Ref. 20).

As discussed, MnAs downscaling results in enhancement of the on-site *d-d* Coulomb repulsion energy *U*. Hence, for decreasing size, hexagonal-MnAs clusters embedded in GaAs approach the boundary, in the (modified) ZSA phase diagram,[20] that separates the regions of *pd*-metals and negative-Δ insulators (arrow 1 in Fig. 3). For decreasing cluster size, the region of negative-Δ insulation widens, due to *p*-bandwidth narrowing. On the other hand, MnAs clusters embedded in GaAs are known to sustain internal strain, leading to enlarged Mn-As bond distance.[22,23] This results in a weakening of the local *p-d* hybridization that counteracts gap opening. Hence, due to the internal nanocluster strain, the region of negative-Δ insulation shrinks (arrow 2 in Fig. 3), and the nanoclusters extend their metallic character somewhat. Sufficiently large clusters are metallic. However, a metal-insulator (metal-semiconductor) transition, involving the opening of a gap of *p-p* type, is expected to eventually occur for decreasing cluster size. Nevertheless, the hexagonal-MnAs crystal structure might become unstable below a certain critical size.

### D. Metal-insulator transition in $Ga_{1-x}Mn_xAs$

The energy levels introduced by a substitutional Mn impurity in GaAs can be thought as the result of the hybridization between the As dangling bonds generated by a Ga vacancy, and the crystal-field and exchange split *d* levels of a $Mn^{3+}$ ion placed at the vacant site.[12] As discussed, the $Mn^{3+}$ oxidation state corresponds to the "ionic limit" for the Mn-As bond (i.e., *p-d* hybridization switched off). In such a limit, the order and filling of the ($Mn^{3+}$)-levels is: $e_+^2$, $t_+^2$, $e_-^0$, $t_-^0$ (Fig. 11 in Ref. 12). Here, $e_+$ and $e_-$ ($t_+$ and $t_-$) denote two-fold (three-fold) degenerated spin-up and spin-down states, respectively, with *e* ($t_2$) symmetry. The configuration is $d^4$, with one ($t_+$) hole (localized in the Mn atom) relative to the $d^5$ configuration of the isolated (neutral) Mn atom. Switching on *p-d* hybridization, two pairs of states with $t_2$ symmetry are generated (Ref. 12): one localized primarily on the Mn atom, referred to as crystal-field resonances, $t_{CFR+}$ and $t_{CFR-}$, and the other localized primarily on the neighboring As atoms, referred to as dangling-bond hybrids, $t_{DBH+}$ and $t_{DBH-}$. In addition, a pair of nonbonding states with *e* symmetry, localized on the Mn atom, $e_{CFR+}$ and $e_{CFR-}$, is also introduced. The order and filling of levels becomes: $t_{CFR+}^3$, $e_{CFR+}^2$, $t_{DBH-}^3$, $t_{DBH+}^2$, $e_{CFR-}^0$, $t_{CFR-}^0$ (Fig. 11 in Ref. 12). While the $t_{DBH+}$ level lies in the band gap, close to the valence-band maximum, the other occupied levels ($t_{CFR+}$, $e_{CFR+}$, and $t_{DBH-}$) lie within the valence band.[12] The configuration is $d^5\underline{L}$, that is, $d^5$ on the Mn atom ($t_{CFR+}^3$, $e_{CFR+}^2$) plus one ($t_{DBH+}$) hole, localized in the As dangling-bond manifold ($t_{DBH-}^3$, $t_{DBH+}^2$). *p-d* hybridization (covalency of the Mn-As bond) thus leads to migration of the hole from the Mn atom into its four As neighbors. By this mechanism, Mn acts as an acceptor in GaAs. This picture, obtained from first-principle electronic-structure calculations,[12] is qualitatively consistent with the results of the ($Mn^{3+}$) CI analysis of the $Ga_{1-x}Mn_xAs$ Mn 2*p* photoemission.[30] The spin-up dangling-bond-hybrid level, $t_{DBH+}$, where the hole resides, is the origin of the so-called Mn-derived impurity band in $Ga_{1-x}Mn_xAs$. It corresponds to the *localized* state that splits off the continuum, for sufficiently strong *p-d* hybridization, in negative-$\Delta$ compounds.[31]

The on-site *d-d* Coulomb repulsion energy $U$ is found to be similar for embedded MnAs nanoclusters, $U = 3.6$ eV, and for the $Ga_{0.93}Mn_{0.07}As$ diluted magnetic semiconductor, $U = 3.5$ eV (Ref. 30), indicating that the degree of localization of *d*-electrons is similar in both materials. Homogeneous $Ga_{1-x}Mn_xAs$ is known to be metallic or nearly metallic. It is expected to be located, in the ZSA phase diagram, close to the boundary that separates the regions of *pd*-metals and negative-$\Delta$ insulators. As a negative-$\Delta$ material, the metal-insulator transition in $Ga_{1-x}Mn_xAs$ is controlled by the lattice strain (affecting the local *p-d* hybridization), as well as by the distribution of Mn atoms (affecting the efficiency of extended Mn-As-Mn bonding interactions). Since Mn atoms are distant from each other and the cation-As-cation bond angle is close to 90°, extended Mn-As-Mn bonding interactions are hardly efficient in $Ga_{1-x}Mn_xAs$.

Hence, the ($t_{DBH+}$)-state has a chance to survive as a *split-off* state. That is, the impurity band has a chance to remain separated from the valence-band edge (by a gap of *p-p* type), even for relatively high Mn concentration. On the other hand, substitutional incorporation of Mn in GaAs is known to lead to notable expansion of the lattice, despite the Mn atomic radius is smaller than that of Ga. This possibly arises on an energetic advantage for a weakened Mn 3*d*-As 4*p* hybridization, making the impurity band to approach the valence-band edge, favoring gap closure and long-range ferromagnetism. Hence, although $Ga_{1-x}Mn_{x}As$ approaches the metal-insulator boundary from the insulator side, increased lattice strain weakens the local *p-d* hybridization, driving $Ga_{1-x}Mn_{x}As$ very close to (or into) the metallic state. In this context, the conduction properties of $Ga_{1-x}Mn_{x}As$ are expected to largely vary, depending on the Mn-content (responsible for the degree of strain) and on the Mn-distribution (affecting extended Mn-As-Mn interactions).

Modifications of the homogeneous Mn distribution upon annealing are expected to alter the degree of localization of the holes mediating ferromagnetism. Spinodal decomposition is known to involve a change in the sign of the lattice strain.[3] Whereas the lattice of homogeneous $Ga_{1-x}Mn_{x}As$ expands relative to that of GaAs, the lattice of spinodal-decomposed $Ga_{1-x}Mn_{x}As$ slightly contracts. The later strengthens the local *p-d* hybridization, such that the region of negative-$\Delta$ insulation widens (arrow 3 in Fig. 3). For low Mn content, strengthening of the local *p-d* hybridization upon spinodal decomposition is expected to favor gap opening, i.e., the insulator regime. However, for sufficiently high Mn content, formation of connecting Mn-As-Mn paths in late stages of spinodal decomposition might favor extended Mn-As-Mn bonding interactions, resulting in a broadening of the impurity band and favoring gap closure. The effect of spinodal decomposition is thus expected to depend on the Mn content, on whether percolation is achieved or not.

# V – SUMMARY

We have carried out a configuration-interaction (CI) cluster-model analysis of Mn 2$p$ photoemission from MnAs nanoclusters embedded in GaAs (GaAs:MnAs). The analysis has revealed that the nanoclusters belongs to the special class of materials with *negative* charge-transfer energy $\Delta$. Homogeneous $Ga_{1-x}Mn_xAs$ appears to belong to this class too. We have made a rational connection between the electronic properties of homogeneous $Ga_{1-x}Mn_xAs$ and phase-separated GaAs:MnAs materials. For both, the metal-insulator (metal-semiconductor) transition is driven by the opening of a gap of $p$-$p$ type (covalent gap). MnAs clusters approach the metal-insulator boundary from the metal side. They are expected to eventually enter the insulator (semiconductor) regime for decreasing size. Homogeneous $Ga_{1-x}Mn_xAs$ approaches the metal-insulator boundary from the insulator side. Strain, leading to weakened $p$-$d$ hybridization, brings $Ga_{1-x}Mn_xAs$ close to (or into) the metal side.

# ACKNOWLEDGMENTS

This work has been partly supported by the Spanish Ministry of Science and Innovation ("Ramón y Cajal" program and MAT2007–66719 grant).

**FIGURE CAPTIONS**

**Figure 1.** (Color online) Lineshape analysis of the Mn 2*p* photoemission spectrum recorded under 750-eV light excitation for MnAs nanoclusters embedded in GaAs (black dots), after subtraction of the secondary-electron background. The vertical black (blue) bars are the calculated unbroadened spectral lines corresponding to elastic photoemission, either main (m) or satellite (s1, s2) contributions. The vertical grey (green) bars are unbroadened spectral lines corresponding to the excitation of bulk ($p_b$) or interface ($p_i$) plasmons. The corresponding broadened spectral contributions are shown by dotted black (blue) curves (elastic photoemission) and dash-dotted grey (green) curves (plasmon losses). The solid grey curve represents the theoretical spectrum obtained adding up all contributions.

**Figure 2.** Energy-level diagram for MnAs nanoclusters embedded in GaAs in the $MnAs_6$ cluster model.

**Figure 3.** Sketch of the location of bulk-MnAs (open dot), phase-separated GaAs:MnAs (solid black dot), and homogeneous $Ga_{1-x}Mn_xAs$ (solid grey diamond) in the (modified) Zaanen-Sawatzky-Allen diagram (Ref. 20) as a function of the charge-transfer energy ($\Delta$) and on-site *d-d* Coulomb repulsion (*U*). The black dotted (solid) line indicates the metal-insulator boundary for unstrained (strained) hexagonal-MnAs. Arrow 2 indicates the shift of the metal-insulator boundary caused by strain. The grey solid (dotted) line indicates the metal-insulator boundary for homogeneous (spinodal decomposed) $Ga_{1-x}Mn_xAs$. Arrow 3 indicates the shift of the metal-insulator boundary caused by strengthened local *p-d* hybridization upon spinodal decomposition. Arrow 1 indicates the enhancement of *U* that occurs upon MnAs downscaling.

**TABLE I.** Electronic structure parameters for MnAs nanoclusters embedded in GaAs (GaAs:MnAs), for selected binary manganese compounds, and for selected diluted magnetic semiconductors. $\Delta$, $\Delta_{\text{eff}}$, $U$, $U_{\text{eff}}$, and $(pd\sigma)$ are given in units of eV.

| | Nominal charges | Nominal ionic configuration | Dominant configuration | $\Delta$ | $\Delta_{\text{eff}}$ | $U$ | $U_{\text{eff}}$ | $(pd\sigma)$ | |
|---|---|---|---|---|---|---|---|---|---|
| MnF$_2$ | Mn$^{2+}$F$_2^{1-}$ | $d^5$ | $d^5$ | 9.0 | | 3.2 | | | Ref. 38 |
| MnCl$_2$ | Mn$^{2+}$Cl$_2^{1-}$ | $d^5$ | $d^5$ | 4.5 | | 3.2 | | | Ref. 38 |
| MnBr$_2$ | Mn$^{2+}$Br$_2^{1-}$ | $d^5$ | $d^5$ | 3.2 | | 3.2 | | | Ref. 38 |
| MnS | Mn$^{2+}$S$^{2-}$ | $d^5$ | $d^5$ | 1.5 | | 4 | | −1.2 | Ref. 33 |
| MnSe | Mn$^{2+}$Se$^{2-}$ | $d^5$ | $d^5$ | 0.8 | | 4 | | −1.0 | Ref. 33 |
| MnTe | Mn$^{2+}$Te$^{2-}$ | $d^5$ | | 0 | | 4 | | −0.8 | Ref. 33 |
| MnAs | Mn$^{3+}$As$^{3-}$ | $d^4$ | | | | | | | |
| GaAs:MnAs | Mn$^{3+}$As$^{3-}$ | $d^4$ | $d^5\underline{L}$ | −2.1 | −4.7 | 3.6 | 3.0 | −0.97 | This work |
| Cd$_{1-x}$Mn$_x$S | Cd$_{1-x}^{2+}$Mn$_x^{2+}$S$^{2-}$ | $d^5$ | $d^5$ | 3.0 | | 4.0 | | −1.3 | Ref. 34 |
| Cd$_{1-x}$Mn$_x$Se | Cd$_{1-x}^{2+}$Mn$_x^{2+}$Se$^{2-}$ | $d^5$ | $d^5$ | 2.5 | | 4.0 | | −1.2 | Ref. 34 |
| Cd$_{1-x}$Mn$_x$Te | Cd$_{1-x}^{2+}$Mn$_x^{2+}$Te$^{2-}$ | $d^5$ | $d^5$ | 2.0 | | 4.0 | | −1.1 | Ref. 34 |
| Ga$_{1-x}$Mn$_x$As | Ga$_{1-x}^{3+}$Mn$_x^{3+}$As$^{3-}$ | $d^4$ | $d^5\underline{L}$ | −1.5 | | 3.5 | | −1.0 | Ref. 30 |

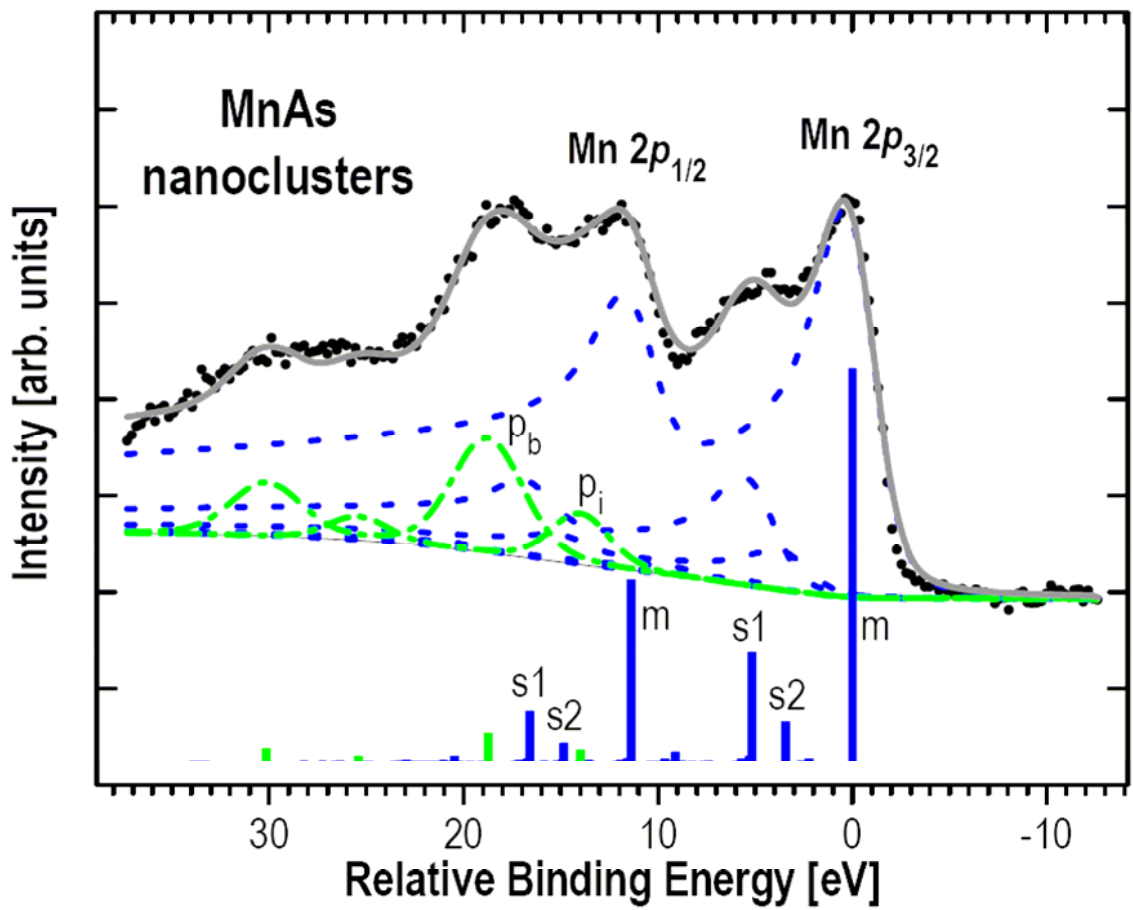

**Figure 1**

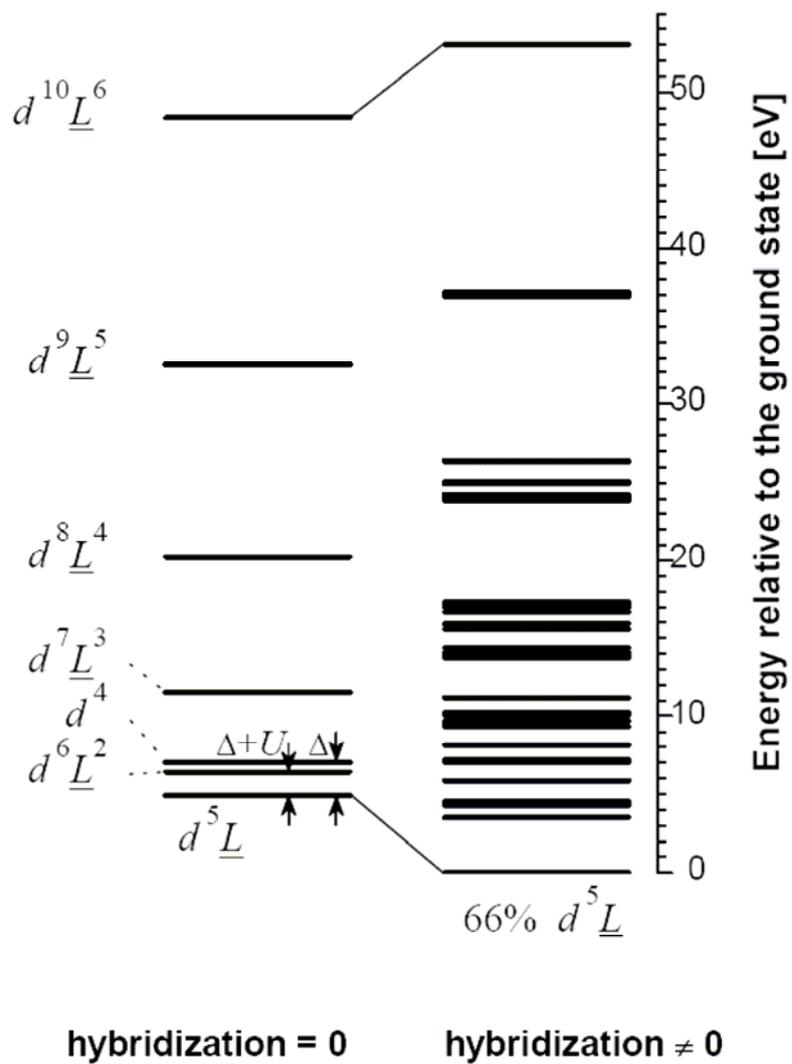

**Figure 2**

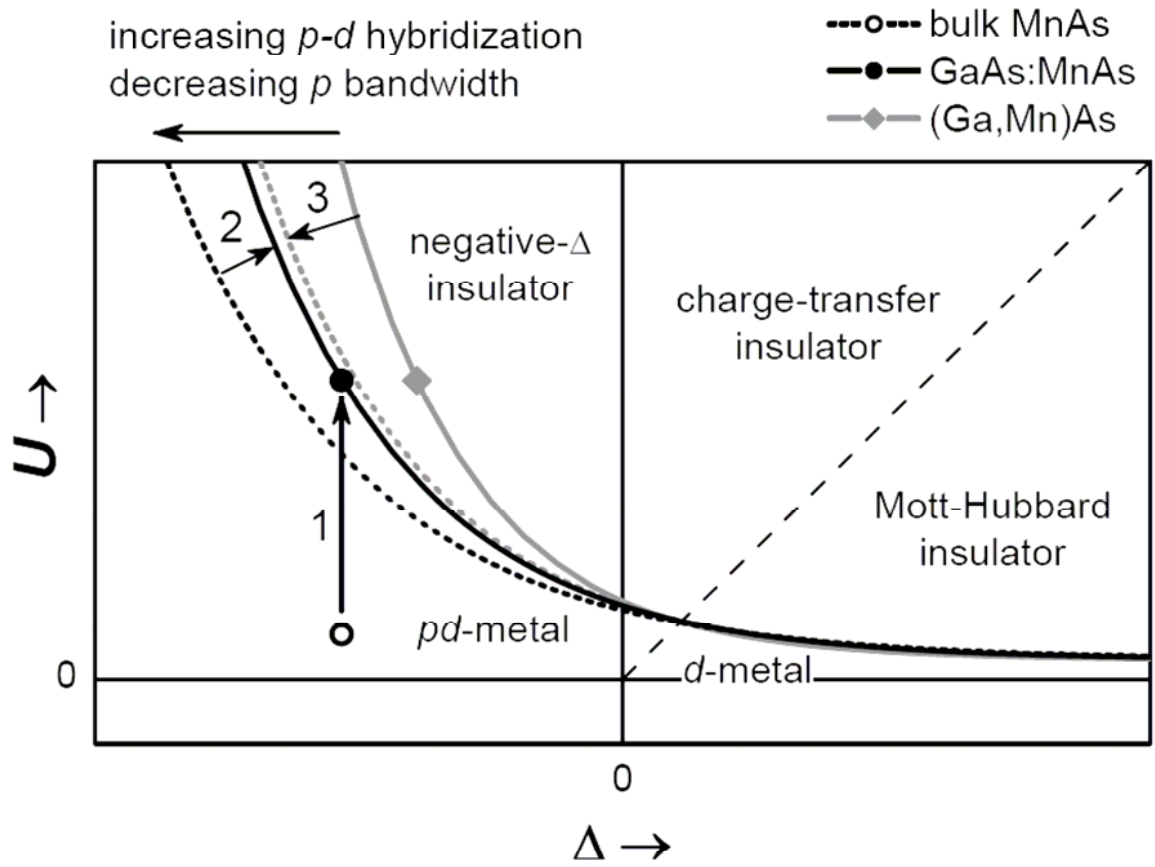

**Figure 3**